\documentclass[a4paper,10.5pt,twoside]{article}
\usepackage[top=3cm, bottom=3cm, inner=3cm, outer=3cm, includehead]{geometry}
\usepackage{fancyhdr}
\usepackage{xurl}
\usepackage{graphicx}
\usepackage{alltt}
\usepackage{amsmath}
\usepackage[hidelinks]{hyperref}
\usepackage[T1]{fontenc}
\usepackage{lmodern}
\usepackage{booktabs} 
\usepackage{csquotes}
\usepackage[backend=biber, style=numeric, sorting=none]{biblatex}
\bibliography{references}
\usepackage[english]{babel}

\begin{document}
\fancyhead[LE]{\thepage\ \ \ \ {exalsius research}}
\fancyhead[RO]{{exalsius research}\ \ \ \ \thepage}
\begin{center}
\LARGE
\textbf{What happens when nanochat meets DiLoCo?}\\[12pt]
\normalsize
\textbf {Alexander Acker}\textsuperscript{1}, 
\textbf {Sören Becker}\textsuperscript{1},
\textbf {Sasho Nedelkoski}\textsuperscript{2},
\textbf {Dominik Scheinert}\textsuperscript{1},
\textbf {Odej Kao}\textsuperscript{3}, 
\textbf {Philipp Wiesner}\textsuperscript{1} \\
\small
\textsuperscript{1}{Team exalsius, Germany} \\
\textsuperscript{2}Independent Researcher, Germany \\
\textsuperscript{3}Technische Universität Berlin, Germany \\
\textit{Authors are listed in alphabetical order.}
\end{center}

\begin{abstract}
Although LLM training is typically centralized with high-bandwidth interconnects and large compute budgets, emerging methods target communication-constrained training in distributed environments. The model trade-offs introduced by this shift remain underexplored, and our goal is to study them.

We use the open-source nanochat project~\cite{nanochat}, a compact 8K-line full-stack ChatGPT-like implementation containing tokenization, pretraining, fine-tuning, and serving, as a controlled baseline. We implement the DiLoCo algorithm~\cite{douillard2024dilocodistributedlowcommunicationtraining} as a lightweight wrapper over nanochat's training loop, performing multiple local steps per worker before synchronization with an outer optimizer, effectively reducing communication by orders of magnitude. This inner-outer training is compared against a standard data-parallel (DDP) setup. Because nanochat is small and inspectable, it enables controlled pipeline adaptations and allows direct comparison with the conventional centralized baseline.

DiLoCo achieves stable convergence and competitive loss in pretraining but yields worse MMLU, GSM8K, and HumanEval scores after mid-training and SFT. We discover that using DiLoCo-pretrained weights and running mid- and post-training with DDP fails to recover performance, revealing irreversible representation drift from asynchronous updates that impairs downstream alignment. We provide this implementation as an official fork of nanochat on GitHub~\footnote{https://github.com/exalsius/nanochat-diloco}.
\end{abstract}

\section{Introduction}
Training large language models (LLMs) has significantly advanced natural language processing. Today, their training remains mostly limited to centralized compute environments due to the demand for high-bandwidth interconnects and substantial computational resources. This reliance comes from the communication-intensive nature of synchronous distributed optimization, such as standard data-parallel training (e.g., Distributed Data Parallelism, or DDP), where gradients are synchronized frequently across accelerators. However, these requirements limit the training of LLMs in settings with limited bandwidth, heterogeneous hardware, or geographically dispersed resources.

An emerging paradigm of distributed low-communication training methods seeks to mitigate these constraints by decoupling local optimization from global synchronization, enabling LLMs to be trained in more flexible and resource-constrained distributed setups. This shift promises to unlock opportunities beyond traditional tightly coupled centralized clusters. Within a datacenter, it relaxes replica placement constraints and improves overall resource utilization. Across independent small clusters, it enables federated multi-organization training, making large-scale models feasible without proprietary high-speed interconnects. And at the edge, it supports on-device fine-tuning that leverages local data with minimal transfer and stronger privacy.

Despite these promises, the trade-offs introduced by this shift toward distributed low-communication training, particularly in terms of model quality, convergence stability, and downstream adaptability remain underexplored. Our work aims to investigate these implications, providing insights into the viability of low-communication approaches for end-to-end LLM training. Prior work has empirically demonstrated convergence under certain assumptions, but evaluations in full LLM pipelines are underexplored, often overlooking how these dynamics influence mid- and post-training stages. In this context, our study explores these trade-offs through controlled experiments.

As a foundation for this investigation, we build upon nanochat~\cite{nanochat}, an open-source project that delivers a minimal yet comprehensive implementation of an LLM system. Encompassing tokenization, pretraining, fine-tuning, and serving within approximately 8,000 lines of code, nanochat serves as an accessible research baseline for dissecting distributed training mechanics, communication patterns, and optimizer behaviors. Its compact design, training a reference 550M-parameter model on 8 H100-class GPUs in a single node, allows for rapid iteration and end-to-end experimentation without the complexity of larger frameworks.

We integrate the DiLoCo algorithm~\cite{douillard2024dilocodistributedlowcommunicationtraining} into nanochat's pretraining and fine-tuning pipelines, enabling distributed low-communication training across workers. DiLoCo decouples optimization by allowing each worker to execute multiple local optimizer steps before periodically sharing parameter deltas, which are aggregated through a global momentum-based optimizer. This approach mirrors the convergence of fully synchronous training while reducing communication frequency. By implementing DiLoCo as a wrapper over nanochat's training loop, we conduct direct comparisons against the standard DDP baseline. Our analysis uncovers strong alignment between DiLoCo-reported results and the baseline during pretraining, but a notable collapse in downstream performance on tasks such as MMLU, GSM8K, and HumanEval following dialogue mid-training and supervised fine-tuning (SFT). Even when DiLoCo-pretrained weights are subsequently trained under DDP, the deficits persist, empirically demonstrating an irreversible representation drift induced by asynchronous updates that compromises the model's downstream alignment capacity. These findings underscore critical limitations in low-communication distributed training and reveal the necessity for improved strategies that mitigate them in future decentralized LLM systems.

\section{Background}
\subsection{nanochat}
nanochat~\cite{nanochat} is a compact, open-source, full-stack implementation of a ChatGPT-like LLM training pipeline, released by Andrej Karpathy. It is designed as a pedagogical and hackable baseline implementation that includes the entire LLM lifecycle, from tokenization to serving, in a minimal, dependency-light codebase of approximately 8,000 lines, emphasizing reproducibility and accessibility for researchers and practitioners. This enables end-to-end training and deployment on modest hardware, such as a single $8\times$ H100 GPU node, at a cost of around \$100 and in under 4 hours, prioritizing transparency over scale to demystify modern LLM development. Specifically, nanochat implements the following key stages of a ChatGPT-style training pipeline:

\begin{itemize}
\item Tokenization: A custom Byte-Pair Encoding (BPE) tokenizer trained from scratch, implemented in Rust for high-speed, memory-efficient text processing.

\item Architecture: A decoder-only transformer including multi-head self-attention, feed-forward MLP blocks, residual connections, and layer normalization, with a reference configuration of 20 layers and $\sim 550M$ parameters.

\item Training pipeline: Pretraining on a shuffled subset of FineWeb-Edu (a high-quality educational corpus), mid-training on dialogue datasets such as SmolTalk augmented with auxiliary tasks (e.g., MMLU and GSM8K with tool-use tags), supervised fine-tuning (SFT) on ARC-Easy/Challenge, GSM8K, and SmolTalk, and optional simplified reward-model-free preference optimization (GRPO) on GSM8K.

\item Evaluation and serving: Built-in evaluation on metrics such as CORE, ARC, MMLU, GSM8K, and HumanEval, alongside a lightweight inference engine (with KV cache and Python interpreter support) and a minimal web UI for interactive chatting, deployable via a single endpoint.
\end{itemize}

This reference setup reproduces LLM training in a transparent manner, yielding a model that outperforms GPT-2 on basic tasks while fitting within moderate hardware budgets. Its inspectable structure makes nanochat a strong starting point for experimentation with distributed optimization methods such as low-communication training paradigms.

\subsection{Distributed Low-Communication Optimization (DiLoCo)}
Distributed Low-Communication (DiLoCo) optimization~\cite{douillard2024dilocodistributedlowcommunicationtraining} is a data-parallel training algorithm designed to relax the communication bottlenecks inherent in synchronous methods such as DDP. By enabling LLM training across loosely connected devices, including geographically dispersed clusters with limited bandwidth, DiLoCo promises scalable distributed optimization with minimal impact on model quality. It draws inspiration from federated averaging but scales the number of inner optimization steps ($H$) to the hundreds, drastically reducing synchronization frequency while empirically maintaining stable convergence.

Recent extensions, such as scaling laws~\cite{charles2025communicationefficientlanguagemodeltraining}, asynchronous variants~\cite{douillard2025streaming}, and open-source implementations like OpenDiLoCo~\cite{jaghouar2024opendiloco}, have demonstrated general convergence during pretraining for models up to 10B parameters, achieving up to $500\times$ communication savings.

The core mechanism decouples local computation from global synchronization. Each of $k$ workers receives a shared model copy $\theta_t$ and performs $H$ inner optimization steps on its local data using a high-fidelity optimizer (typically AdamW or, more recently, Muon), yielding a parameter delta $\Delta\theta_i = \theta_i^H - \theta_t$. These deltas are then averaged,
\[
\bar{\Delta\theta} = \frac{1}{k} \sum_i \Delta\theta_i,
\]
and applied via an outer optimizer, typically SGD with Nesterov momentum ($\mu > 0$):
\[
v_{t+1} = \mu v_t + \bar{\Delta\theta}, \quad
\theta_{t+1} = \theta_t + \eta v_{t+1},
\]
where $v_t$ is the momentum buffer and $\eta$ is the outer learning rate. This outer step aims to preserve generalization benefits close to full-batch SGD, while the inner steps amortize communication costs. Empirical results demonstrate that DiLoCo matches DDP perplexity on C4 during pretraining using up to 8 model replicas, while requiring significantly less communication and tolerating latencies of several seconds per synchronization. However, current literature focuses primarily on matching loss and perplexity between DDP and DiLoCo; benchmark performance, particularly downstream task performance after fine-tuning, remains underexplored.

\section{Training nanochat with DiLoCo}
We extended the nanochat training loop to include a distributed DiLoCo wrapper. Each worker executes $H$ local AdamW and Muon optimizer steps and periodically synchronizes pseudo-gradients (parameter deltas). The outer optimization is performed using Nesterov momentum SGD on the averaged deltas. Key hyperparameters were set as follows:

\begin{itemize}
\item Inner optimizers: AdamW and Muon (default in nanochat), used in non-DDP mode.
\item Outer optimizer: SGD with Nesterov momentum ($\mu_{\text{outer}} = 0.9$).
\item Synchronization interval ($H$): $H = 100$ for base pretraining and $H = 30$ for mid-training and SFT.
\item Outer learning rate: $\eta_{\text{outer}} = 0.8$.
\item Global batch size: $2^{19}$ tokens across 8 NVIDIA A100 GPUs.
\end{itemize}

The synchronization interval and hyperparameters were chosen based on related scaling-law studies~\cite{charles2025communicationefficientlanguagemodeltraining} for $M = 8$ workers and a $\sim550M$ - parameter model.

\section{Evaluation and Results}

\subsection{Benchmark comparison}
We evaluate three configurations: (1) Standard DDP (fully synchronous as in the publicly released nanochat pipeline), (2) DiLoCo (low-communication across all stages), and (3) Hybrid (DiLoCo-pretrained base, followed by DDP mid-training and SFT).

\begin{table}[h!]
\centering
\caption{Comparison across Standard DDP, DiLoCo, and Hybrid. CORE is reported for the base stage; mid-training and SFT show benchmark validation metrics.}
\label{tab:joint_results_updated}
\resizebox{\linewidth}{!}{
\begin{tabular}{lcccccccc}
\toprule
\textbf{Stage} & \textbf{Method} & \textbf{CORE} & \textbf{ARC-Easy} & \textbf{ARC-Chal.} & \textbf{MMLU} & \textbf{GSM8K} & \textbf{HumanEval} & \textbf{ChatCORE} \\
\midrule
Base & Standard DDP & \textbf{0.2065} & -- & -- & -- & -- & -- & -- \\
Base & DiLoCo & 0.1788 & -- & -- & -- & -- & -- & -- \\
Base & Hybrid & -- & -- & -- & -- & -- & -- & -- \\
\midrule
Mid & Standard DDP & -- & \textbf{0.4133} & \textbf{0.2969} & \textbf{0.3301} & \textbf{0.0394} & \textbf{0.0710} & \textbf{0.1060} \\
Mid & DiLoCo & -- & 0.2462 & 0.2619 & 0.2538 & 0.0190 & 0.0610 & 0.0192 \\
Mid & Hybrid & -- & 0.2475 & 0.2611 & 0.2509 & 0.0212 & 0.0488 & 0.0165 \\
\midrule
SFT & Standard DDP & -- & \textbf{0.4356} & \textbf{0.3200} & \textbf{0.3300} & \textbf{0.0561} & \textbf{0.0740} & \textbf{0.1190} \\
SFT & DiLoCo & -- & 0.2395 & 0.2432 & 0.2469 & 0.0447 & 0.0610 & 0.0157 \\
SFT & Hybrid & -- & 0.2466 & 0.2602 & 0.2451 & 0.0311 & 0.0366 & 0.0140 \\
\bottomrule
\end{tabular}}
\end{table}

\paragraph{Overall trends.}
The results in Table~\ref{tab:joint_results_updated} highlight a clear gap between the efficiency and exactness of DiLoCo training. While DiLoCo achieves stable pretraining convergence (CORE of $0.1788$ compared to $0.2065$ for the standard DDP baseline), downstream adaptation collapses across almost all evaluated benchmarks. The Hybrid configuration, which switches to standard DDP after DiLoCo pretraining, fails to recover performance, indicating that the representational drift induced during the DiLoCo phase persists into later stages.

\paragraph{Pretraining stability.}
DiLoCo's base-stage CORE of $0.1788$ is marginally lower than the standard baseline, yet it exhibits smooth and consistent loss trajectories throughout pretraining (Figure~\ref{fig:pretraining_loss}). This confirms that local optimizer accumulation with delayed synchronization can closely approximate synchronous training for next-token prediction. Communication volume was reduced by approximately $100\times$ with near-linear hardware scaling, validating DiLoCos efficiency in bandwidth-constrained environments. These findings align closely with prior reports on DiLoCo and its scaling behavior~\cite{douillard2024dilocodistributedlowcommunicationtraining, charles2025communicationefficientlanguagemodeltraining}. As in those studies, the loss difference of roughly 0.15-0.20 for a $\sim550M$ model appears substantial and limits the effectiveness of mid-training and SFT on dialogue-like data.

\paragraph{Mid-training degradation.}
Divergence becomes pronounced during mid-training on dialogue-centric data. On ChatCORE, DiLoCo underperforms the standard setup by more than $5\times$ ($0.0192$ vs.\ $0.1060$), with similar drops on reasoning tasks such as GSM8K and MMLU. The Hybrid variant, despite returning to DDP, performs similarly poorly ($0.0165$ on ChatCORE), suggesting that DiLoCo pretraining alters internal feature geometry or layer statistics in ways that mid-training cannot correct.

\paragraph{Supervised fine-tuning (SFT).}
The performance gap widens further during SFT. The standard DDP pipeline shows consistent gains (e.g., ChatCORE = $0.1190$; GSM8K = $0.0561$), while both DiLoCo and Hybrid models plateau at or below $0.0157$. This persistent underperformance suggests that DiLoCo induces weight-space distortions that downstream training stages cannot repair.

\paragraph{Interpretation.}
These results indicate that while DiLoCo preserves short-horizon loss minimization, it compromises global representational coherence. The loss of cross-worker synchrony-driven by divergent local momentum and optimizer states, appears to guide models into flatter but semantically misaligned minima. This “alignment fragility” hypothesis is supported by prior findings that delayed synchronization can degrade generalization even when training loss remains stable~\cite{hakimi2020tamingmomentumdistributedasynchronous, dandi2021implicitgradientalignmentdistributed}.

\subsection{Analysis of Training Dynamics}

To better understand the downstream degradation observed in Table~\ref{tab:joint_results_updated}, we analyze loss trajectories across training stages for the Standard, DiLoCo, and Hybrid configurations.

\subsubsection{Base pretraining: Stable convergence with higher loss floor}

\begin{figure}[!ht]
    \centering
    \includegraphics[width=0.8\linewidth]{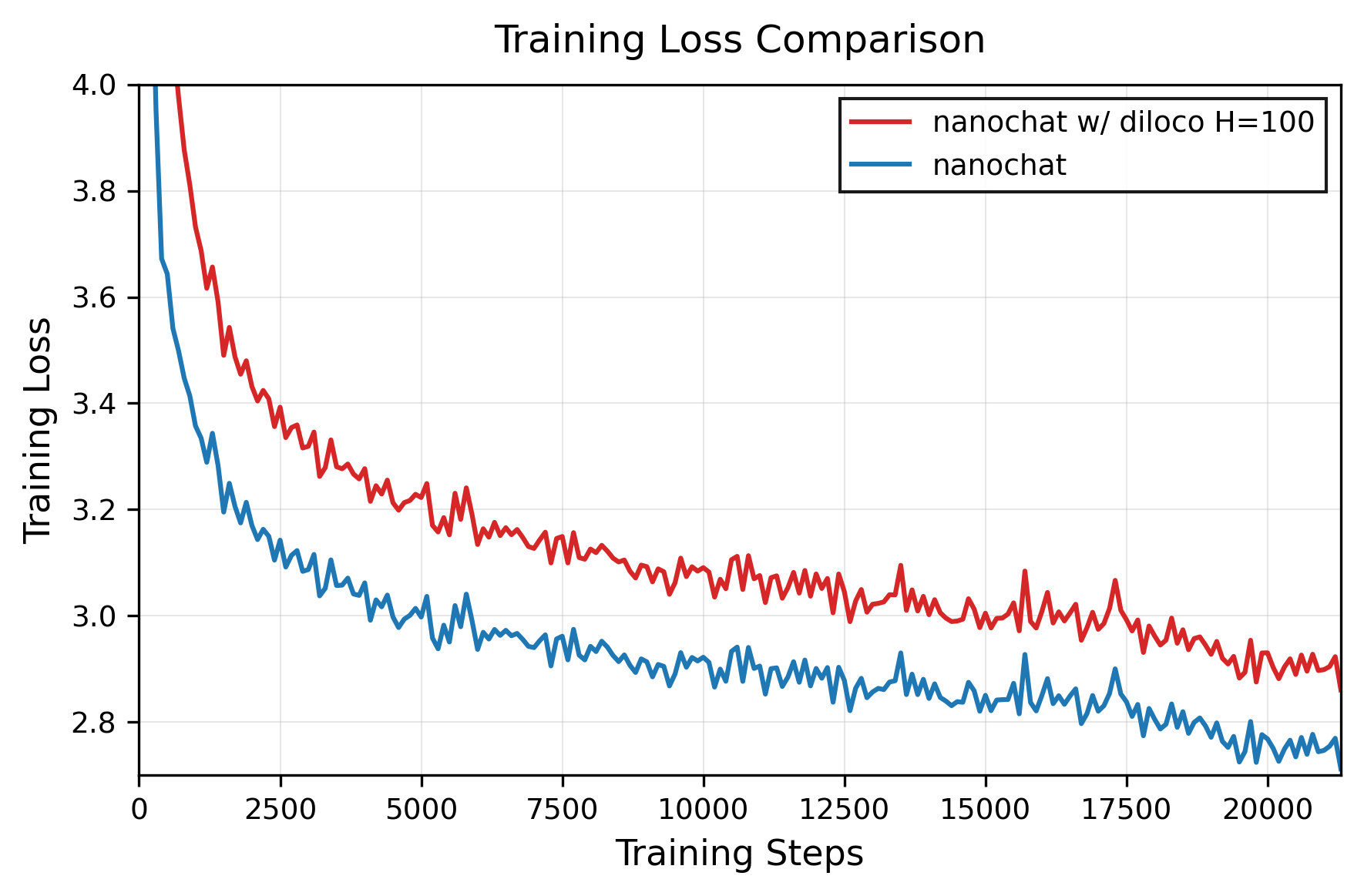}
    \caption{Training loss comparison during the base pretraining stage. DiLoCo (red) achieves stable convergence compared to the Standard (blue) baseline.}
    \label{fig:pretraining_loss}
\end{figure}

During pretraining, both the Standard and DiLoCo models exhibit smooth, monotonic loss curves (Figure~\ref{fig:pretraining_loss}). Both runs use the same number of steps. DiLoCo tracks the baseline trajectory closely, though consistently at a higher absolute loss. We note that, given the chosen settings (model size $\sim 550M$, 8 workers, H100 GPUs), our trends replicate those reported in DiLoCo ablation studies~\cite{charles2025communicationefficientlanguagemodeltraining}. The higher pretraining loss floor appears to impede alignment during mid-training and SFT. As written in ~\cite{charles2025communicationefficientlanguagemodeltraining}, these effects reduce when the model size increases, however, we hypothesise, even then the effects are quite noticable. This is left to be proven empirically.

\subsubsection{Mid-training: Onset of alignment mismatch}

\begin{figure}[!ht]
    \centering
    \includegraphics[width=0.8\linewidth]{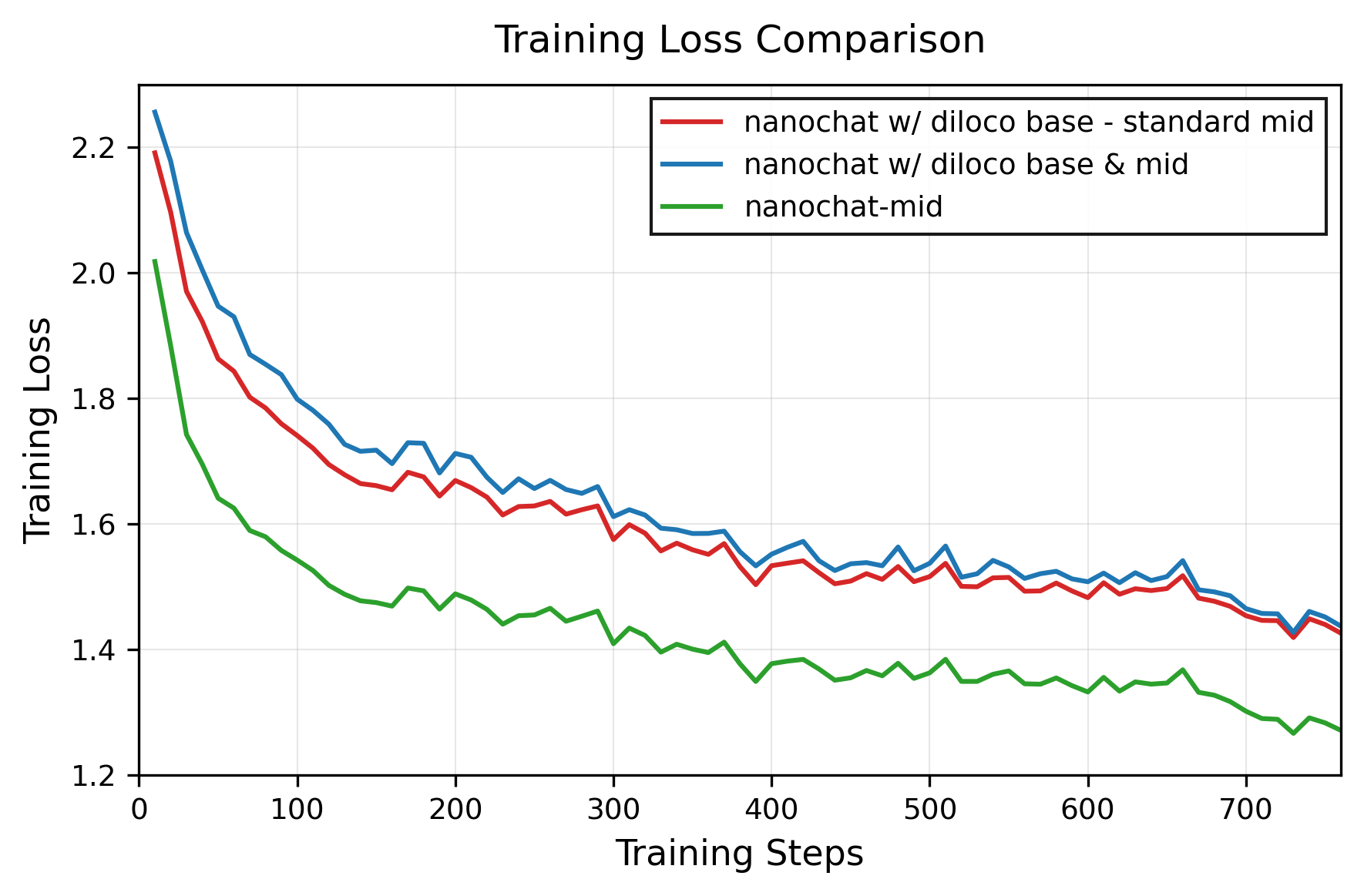}
    \caption{Mid-training loss comparison. DiLoCo (red) and Hybrid (green) models fail to minimize the dialogue-style objective relative to the Standard (blue) baseline.}
    \label{fig:midtraining_loss}
\end{figure}

At mid-training, where the distribution shifts toward dialogue-style instruction data (Figure~\ref{fig:midtraining_loss}), DiLoCo and Hybrid models plateau at significantly higher loss levels, indicating difficulty adapting to the new objective. Even with full synchronization restored, the Hybrid model fails to converge further, suggesting that the representational geometry acquired during DiLoCo pretraining is incompatible with instruction-following objectives.

We hypothesize that this stagnation reflects representation drift: each worker's embedding space diverges during extended local optimization. Though global averaging maintains token-level cross-entropy, the resulting internal states are semantically incoherent, impairing transfer to dialogue tasks.

\subsubsection{SFT: Continued degradation}

\begin{figure}[!ht]
    \centering
    \includegraphics[width=0.8\linewidth]{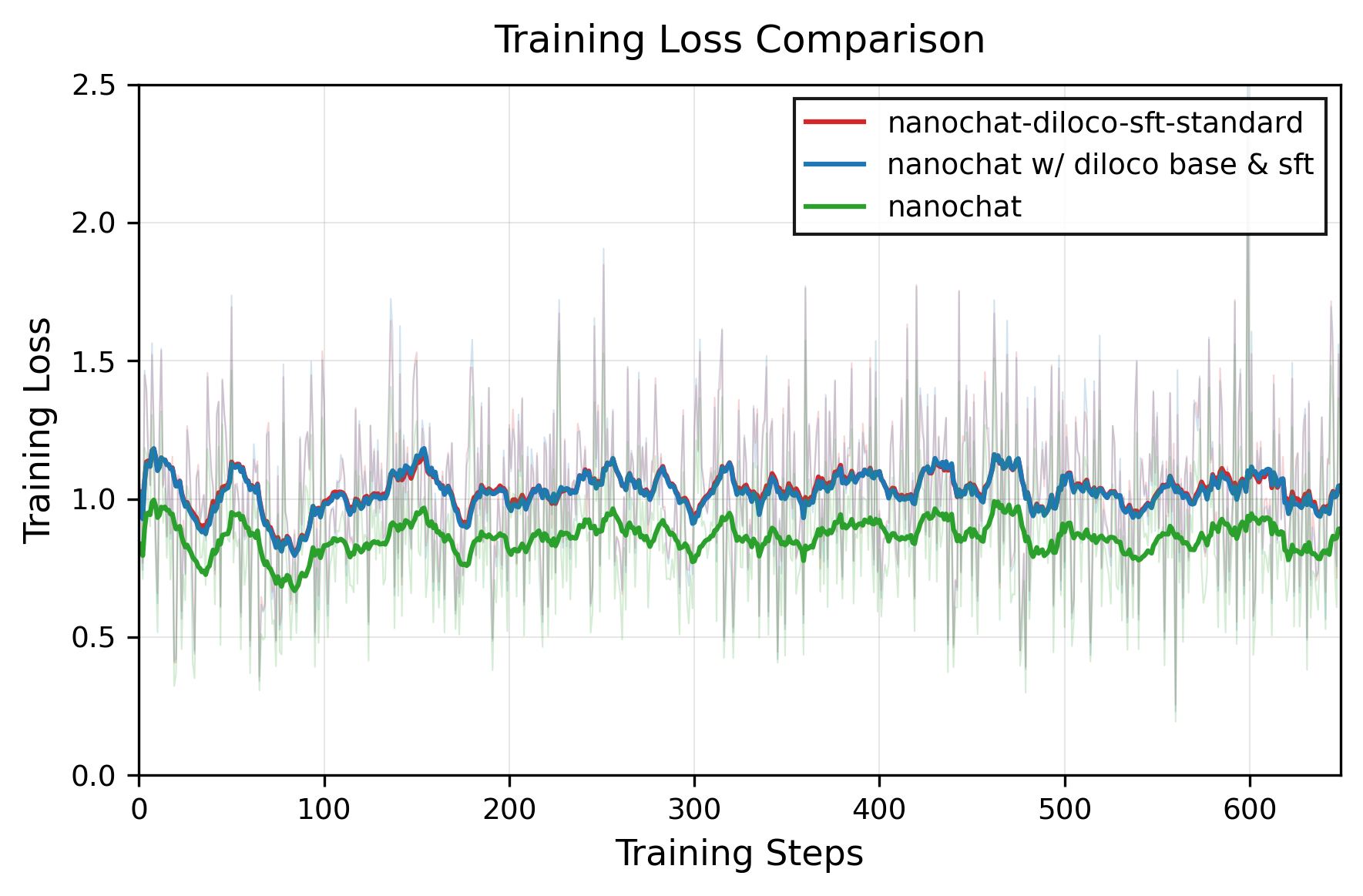}
    \caption{SFT loss comparison. Both DiLoCo (red) and Hybrid (green) exhibit high loss floors, consistent with the quantitative collapse in Table~\ref{tab:joint_results_updated}.}
    \label{fig:sft_loss}
\end{figure}

The degradation intensifies during supervised fine-tuning (Figure~\ref{fig:sft_loss}). The Standard DDP model continues to improve (ChatCORE = $0.1190$; GSM8K = $0.0561$), while DiLoCo and Hybrid models plateau quickly. This stagnation mirrors their near-zero downstream scores, reinforcing that representational damage originating during DiLoCo pretraining is not corrected by later training, even under full synchronization.

\subsection{Possible causes of degradation}
We hypothesize the following mechanisms:

\begin{enumerate}
\item Representation drift: Prolonged local optimization causes workers' embedding spaces to diverge, producing parameter deltas that average into globally coherent but locally inconsistent semantics.

\item Loss-surface bias: Infrequent synchronization steers optimization toward flatter minima that preserve pretraining perplexity but lack the curvature needed for robust transfer to reasoning and dialogue tasks.

\item Alignment degradation: The Hybrid configuration's persistent failure, despite restored DDP synchronization, suggests that DiLoCo-induced representational drift is effectively degraded in later stages.
\end{enumerate}

\section{Conclusion}
This work uses the compact and inspectable nanochat implementation to expose a critical yet understudied trade-off in distributed low-communication training. DiLoCo enables pretraining in low-bandwidth distributed environments, matching synchronous DDP in loss and perplexity, yet it induces degraded representation drift that collapses performance during mid-training and SFT. The persistent failure of Hybrid recovery underscores that asynchronous local updates, while computationally efficient, compromise the representational coherence required for instruction alignment and reasoning.

Despite the reduced downstream accuracy, this integration establishes a reproducible baseline for studying the generalization-communication trade-off within a complete, open-source LLM pipeline. Future work will explore adaptive synchronization strategies, such as dynamically adjusting $H$, reducing it during critical stages (end of base training, mid-training, SFT) and increasing it during stable pretraining, to mitigate drift while preserving communication efficiency. These findings highlight the fragility of downstream generalization under communication sparsity and call for new mitigation strategies, such as drift-aware averaging, to enable decentralized LLM training without sacrificing task fidelity.

\section*{Acknowledgments}
We thank the nanochat maintainers for open-sourcing a research-friendly LLM pipeline.

\printbibliography

@article{jaghouar2024opendiloco,
  title={Opendiloco: An open-source framework for globally distributed low-communication training},
  author={Jaghouar, Sami and Ong, Jack Min and Hagemann, Johannes},
  journal={arXiv preprint arXiv:2407.07852},
  year={2024}
}

@article{douillard2025streaming,
  title={Streaming diloco with overlapping communication: Towards a distributed free lunch},
  author={Douillard, Arthur and Donchev, Yanislav and Rush, Keith and Kale, Satyen and Charles, Zachary and Garrett, Zachary and Teston, Gabriel and Lacey, Dave and McIlroy, Ross and Shen, Jiajun and others},
  journal={arXiv preprint arXiv:2501.18512},
  year={2025}
}

@misc{nanochat,
  author = {Andrej Karpathy},
  title = {nanochat: The best ChatGPT that \$100 can buy},
  year = {2025},
  publisher = {GitHub},
  url = {https://github.com/karpathy/nanochat}
}

@misc{douillard2024dilocodistributedlowcommunicationtraining,
      title={DiLoCo: Distributed Low-Communication Training of Language Models}, 
      author={Arthur Douillard and Qixuan Feng and Andrei A. Rusu and Rachita Chhaparia and Yani Donchev and Adhiguna Kuncoro and Marc'Aurelio Ranzato and Arthur Szlam and Jiajun Shen},
      year={2024},
      eprint={2311.08105},
      archivePrefix={arXiv},
      primaryClass={cs.LG},
      url={https://arxiv.org/abs/2311.08105}, 
}

@misc{charles2025communicationefficientlanguagemodeltraining,
      title={Communication-Efficient Language Model Training Scales Reliably and Robustly: Scaling Laws for DiLoCo}, 
      author={Zachary Charles and Gabriel Teston and Lucio Dery and Keith Rush and Nova Fallen and Zachary Garrett and Arthur Szlam and Arthur Douillard},
      year={2025},
      eprint={2503.09799},
      archivePrefix={arXiv},
      primaryClass={cs.LG},
      url={https://arxiv.org/abs/2503.09799}, 
}

@misc{hakimi2020tamingmomentumdistributedasynchronous,
      title={Taming Momentum in a Distributed Asynchronous Environment}, 
      author={Ido Hakimi and Saar Barkai and Moshe Gabel and Assaf Schuster},
      year={2020},
      eprint={1907.11612},
      archivePrefix={arXiv},
      primaryClass={cs.LG},
      url={https://arxiv.org/abs/1907.11612}, 
}

@misc{dandi2021implicitgradientalignmentdistributed,
      title={Implicit Gradient Alignment in Distributed and Federated Learning}, 
      author={Yatin Dandi and Luis Barba and Martin Jaggi},
      year={2021},
      eprint={2106.13897},
      archivePrefix={arXiv},
      primaryClass={cs.LG},
      url={https://arxiv.org/abs/2106.13897}, 
}
\end{document}